\documentclass[twocolumn,prl,showpacs,preprintnumbers]{revtex4}
\usepackage[applemac]{inputenc}
\usepackage[T1]{fontenc} 
\usepackage{amsmath} 
\usepackage{amsfonts} 
\usepackage{amssymb}
\usepackage{graphicx} 
\usepackage[usenames]{color} 

\usepackage{bbm}
\def\beq{\begin{equation}}
\def\eeq{\end{equation}}
\def\bea{\begin{eqnarray}}
\def\eea{\end{eqnarray}}
\def\ba{\begin{array}}
\def\ea{\end{array}}

\def\haf#1{{{#1}\over 2}}

\def\lb{\left(}
\def\rb{\right)}
\def\lr{\left|}
\def\rr{\right|}
\def\l.{\left.}
\def\r.{\right.}

\def\ra{\rangle}
\def\la{\langle}

\def\part{\partial}

\begin{document}

\preprint{UdeM-GPP-TH-08-168}
\preprint{arXiv:0806.xxxx}
\title{Suppression of quantum tunneling for all spins for easy-axis systems.}
\author{Avinash Khare$^1$}
\email{khare@iopb.res.in}
\author{M. B. Paranjape$^{2,3}$} 
\email{paranj@lps.umontreal.ca}
\affiliation{$^1$Institute of Physics, Sachivalaya Marg, Bhubaneswar, Orissa, India, 751005}
\affiliation{$^2$Groupe de physique des particules, D\'epartement de physique,
Universit\'e de Montr\'eal,
C.P. 6128, succ. centre-ville, Montr\'eal, 
Qu\'ebec, Canada, H3C 3J7 }
\affiliation{$^3$Center for Quantum Space-time, Department of Physics, Sogang University,  Shinsu-dong \#1, Mapo-gu, Seoul, 121-742, Korea}

\begin{abstract}  The semi-classical limit of quantum spin systems corresponds to a dynamical Lagrangian which contains the usual kinetic energy, the couplings and interactions of the spins and an additional, first order kinematical term which corresponds to the Wess-Zumino-Novikov-Witten (WZNW) term for the spin degree of freedom \cite{og}.  It was shown that in the case of the kinetic dynamics determined only by the WZNW term, half odd integer spin systems show a lack of tunneling phenomena whereas integer spin systems are subject to it \cite{l} in the case of potentials with easy-plane easy-axis symmetry.  Here we prove, for the theory with a normal quadratic kinetic term of arbitrary strength or the first order theory with azimuthal symmetry (which is equivalently the so-called easy-axis situation),  that the tunneling is  in fact suppressed for all non-zero values of spin.  This model exemplifies the concept that in the presence of complex Euclidean action, it is necessary to use the ensuing complex critical points  in order to define the quantum (perturbation) theory \cite{ampr}.   In the present example, if we do not do so, exactly the opposite, erroneous  conclusion, that the tunneling is unsuppressed for all spins, is reached.
\end{abstract}


\maketitle

{\it Semi-classical spins -} Semiclassical spin systems are governed by an action of the form
\beq
S= \int dt\lb \haf  I {\partial_t \hat s}\cdot{\partial_t \hat s}-V(\hat s)\rb +\sigma\int d^2x\epsilon^{ij}(\hat s\cdot\partial_i\hat s\times\partial_j\hat s)\label{1}
\eeq
where $\hat s\equiv (\sin\theta\cos\varphi, \sin\theta\sin\varphi,\cos\theta)$ is a three vector of unit norm, representing semi-classically the quantum spin \cite{og}.  This description is valid for large value of the spin, which is given by $\sigma$.  The second term is the so-called Wess-Zumino-Novikov-Witten term  \cite{wznw}, which takes into account the fact that the original quantum spin satisfies the algebra of the rotation group.   $\sigma$ the coefficient of the WZNW term is quantized to be a half integer, $\sigma=N/2$.  The WZNW term is defined by extending the spin configuration into an extra, auxiliary dimension.  The arbitrariness of the way to extend the field configuration into the extra dimension  makes the value of the WZNW term ambiguous, however due to the topological nature of the WZNW term, only discretely so.    Thus quantizing the coefficient allows for the action to be well defined modulo $2\pi$.  Classically this is nonsense, but quantum mechanically, where it is only the exponential of the $i$ times the action that is meaningful, this yields a well defined quantum mechanics.  Bosonic spins correspond to integer values of $\sigma$,  {\it i.e.} $N$ even, while fermionic spins correspond to half odd integer values, {\it i.e.} $N$  odd.  

In this paper we consider the spin system with a second order kinetic term in addition to the WZNW term.  For the case of a potential with easy-axis, azimuthal symmetry, with additionally  a reflection symmetry (along the axis), we prove that in fact the tunneling is  suppressed for both bosonic and fermionic non-zero spin systems.  Easy-axis reflection symmetry means $V(\theta,\phi)\equiv V(\theta)=V(\pi-\theta)$, for example,  the potential  $V(\hat s)\equiv V(\theta,\phi)=\haf 1\gamma \sin^2\theta$.   The potential is assumed to be such that  we have two, identical, degenerate classical ground states, one centered at the north pole while the other at the south pole  and corresponding semi-classically described, {\it perturbative}, quantum ground states.   Normally it is expected that there is quantum mechanical, {\it non-perturbative} tunneling between these ground states, usually making the symmetric combination to be the true ground state while the anti-symmetric combination to be slightly lifted in energy.  We will show both semi-classically and analytically that this tunneling does not occur and the two ground states remain exactly degenerate.   The only case for which tunneling persists is for zero spin but his case is actually out of the physical purview of the present analysis.  Indeed the models considered here describe spin systems only in the large $\sigma$ limit.
 
It was shown in Ref. \cite{l}  for situations where the kinetic dynamics is determined only by the WZNW term and where there is assumed an easy-plane easy axis-symmetry in the potential  giving rise to degenerate classical ground states along one axis,  that  tunneling between these ground states is suppressed for fermionic systems while for bosonic systems it is not.  This conclusion was based on the analysis of the contribution of the WZNW term.  It gives an equal contribution for the two paths corresponding to instantons and anti-instantons in the case of bosonic spins, but an equal and opposite contribution in the case of fermionic spins.  This analysis does not keep the quadratic kinetic term as in  Eq. (\ref{1}).  A standard reason to spurn this term is that for low energy dynamics, the second order term would give a negligible correction to predictions from just the first order term.  However, considering the action based only on the first order WZNW term, as done in Ref. \cite{l}), and applying it to our case of easy-axis symmetry (in fact for a slightly generalized separable potential), we easily reproduce the suppression of tunneling for all spins.   The reason for the suppression is different in the two cases with or without the second order term.  Without the second order term, there simply do not exist any instanton trajectories that would mediate the tunneling.  With the second order term,  we find that even for arbitrarily small coefficient $I$ in Eq. (\ref{1}) that although there is an instanton, the tunneling is suppressed because it has infinite action.

The system considered in this paper is an exemplar of the general situation of complex actions with complex critical points \cite{ampr}, and  one that is physically and phenomenologically relevant.  The Euclidean Feynman path integral with imaginary terms in the action, which occurs in the presence of $t$-violating interactions such as the WZNW term, requires deformation of the path integration contour into the space of complexified field configurations, at least in order to define the perturbation theory.  Perturbation theory is based on the idea of quantizing the Gaussian fluctuations about the classical critical points of the action.  For an action with imaginary terms, the critical point is generally attained only at complex values of the field variables.  The contour of path integration should be deformed so as to pass through the complex critical point  in the direction of steepest descent.  Alternatively, it is proposed that it is adequate to consider only the real part of the action to define the measure of functional integration, that the imaginary part only gives a bounded phase factor which can be integrated against this measure.  Although this is probably correct, it is incorrect to base the perturbation theory about the critical points of the real part of the action.  Indeed, we will show that for the spin system under consideration, expanding about these abridged, real critical points erroneously indicates that there is tunneling, whereas it can be shown analytically that there is in fact none.  

{\it Complex path integral - } Quantum mechanically we can calculate the matrix element, corresponding to the Eudclidean persistence amplitude in an eigenstate of angular position $\lr \theta,\varphi\ra \r.$
\beq
\l.\la \theta,\varphi\rr e^\frac{-\beta\hat H}{\hbar}\lr \theta,\varphi\ra \r.=\int{\cal D}(\theta (\tau),\varphi(\tau))e^{-\frac{1}{\hbar}S_E}\label{2}
\eeq
using the Euclidean path integral.  $S_E$ is the Euclidean action obtained by analytically continuing the Minkowski action of Eq. (\ref{1}) to imaginary time, $t\to -i\tau$.  This yields a real positive definite term coming from the normal, quadratic part of the action, while the WZNW term, because it is $t$-odd, remains imaginary:
\beq
S_E=\int d\tau\lb \haf  I {\partial_\tau \hat s}\cdot{\partial_t
\tau \hat s}+V(\hat s)\rb -i\sigma\int d^2x\epsilon^{ij}(\hat s\cdot\partial_i\hat s\times\partial_j\hat s).\label{3}
\eeq
The matrix element on the LHS of Eq. (\ref{2}) admits the expansion in terms of energy eigenstates $|E_i\rangle$
\beq
\l.\la \theta,\varphi\rr e^\frac{-\beta\hat H}{\hbar}\lr \theta,\varphi\ra \r.=\sum_ie^{-\beta E_i\over\hbar}|\langle \theta, \varphi |E_i\rangle|^2. \label{4}
\eeq
In the case  with two degenerate classical ground states, if there is tunneling, there will arise two low lying energy levels contributing to the expansion in Eq. (\ref{4}),  which can be recovered by evaluating the RHS of Eq. (\ref{2}), in the semi-classical approximation.  In this approximation, the energy splitting is calculated to be proportional to $e^{-S_0/\hbar}$,  the hallmark of a tunneling effect, where $S_0$ is the action of the corresponding instanton.  An instanton is a classical solution of the Euclidean equations of motion  that satisfies the appropriate boundary conditions.  We will show that there exist no finite action instanton solutions to the Euclidean equation of motion coming from the action Eq. (\ref{3}).  Thus we find that the amplitude for tunneling simply vanishes.  

The Euclidean equations of motion, because of the imaginary WZNW term contain $i$ explicitly, and as is generally expected, their solution lies in the space of complex configurations.  In the condensed matter literature, there does not seem to be any aversion to analytically continuing the contour of path integration into the space of complex configurations, see for example \cite{c}.  In the particle physics literature, typically, only the real part of the action is used to determine the critical points.  We emphatically assert that this gives the wrong result in the case at hand.  In principle, if the path integral can be done exactly, it should not matter what point is used as a center point.  However, if the path integral is done only in a Gaussian approximation about a center point, then the true answer can be completely obscured.  This is explicitly seen in the case of monopoles in the Georgi-Glashow model, \cite{hst} and a general analysis in  \cite{ampr}.

{\it Abridged, real critical points - } 
If we put $\sigma=0$, we obtain the action corresponding to only the real part of the action.  Varying to find the critical point, we get 
\bea
I\ddot\theta-I\sin\theta\cos\theta\dot\varphi^2 -\frac{\partial V(\theta)}{\partial\theta}&=&0\label{12}\\
I\sin^2\theta\dot\varphi &=&l\label{13}
\eea
where $l$ is a constant which is obtained as an integral of the equations of motion since $\varphi$ is a cyclical coordinate and where the overdot refers to a derivative with respect to Euclidean time $\tau$. Eq. (\ref{12}) determines $\theta (\tau)$ after replacing for $\dot\varphi$ from Eq. (\ref{13}), which then in turn serves to fix $\varphi (\tau)$. The second term in Eq (\ref{12}) serves as a centrifugal barrier, prohibiting the spin to ever be at (depart from) the north pole unless $l=0$.  Therefore this must be the case for the instanton solution that we search.  Eq. (\ref{12}) then  is integrable, yielding
\beq
\frac{I}{2}\dot\theta^2 -V(\theta)=const. =0.
\eeq
The constant is fixed by the boundary condition that at $\tau=-\infty$, the trajectory starts at the north pole with zero velocity.  Integrating to quadrature, assuming  a reflection symmetric double well potential $V(\theta)=V(\pi-\theta)$ with absolute minima at the poles (normalized to zero) yields 
\beq
\int_{\pi/2}^{\theta(\tau)}d\theta\sqrt\frac{ I} {2V(\theta)}=\tau-\tau_0.
\eeq
The action for the trajectory is  simply calculated to be
\beq
S_0=\int d\tau \lb\frac{I}{2}\dot\theta^2 +V(\theta)\rb=\int_0^\pi d\theta\sqrt{2IV(\theta)}
\eeq
using Eq. (\ref{12}), and which is evidently finite.  The contribution of the WZNW term to such a trajectory is strictly zero, since $\dot\varphi=0$, and hence we are led to the conclusion that there is no suppression of tunneling.  This conclusion, as we will show, is entirely incorrect.  

{\it Full, complex critical points - }  In the case at hand, the critical points of the full action satisfy the equations of motion:
\bea
I\ddot\theta -I\sin\theta\cos\theta\dot\varphi^2-\frac{\partial V(\theta)}{\partial\theta} +i\sigma\sin\theta\dot\varphi&=&0\label{9}\\
I\frac{d}{d\tau}\lb\sin^2\theta\dot\varphi\rb -i\sigma\sin\theta\dot\theta&=&0\label{10}
\eea
Notice the explicit $i$ in the equations, which prohibits a solution with real values for both $\theta$ and $\varphi$.  The second equation integrates immediately, analogous to the conservation of azimuthal angular momentum,
\beq
\frac{d}{d\tau}\lb I\sin^2\theta\dot\varphi +i\sigma\cos\theta\rb=0
\eeq
yielding
\beq
I\sin^2\theta\dot\varphi +i\sigma\cos\theta=il\label{16}
\eeq
where we have anticipated that a solution requires imaginary field variables and replaced $l\to il$.  We note that $\varphi$ is completely imaginary while $\theta$ remains real.
Replacing from Eq. (\ref{16}) into Eq. (\ref{10}) and integrating once yields
\beq
\haf I \dot\theta^2 -\haf 1\frac{(l-\sigma\cos\theta)^2}{I\sin^2\theta}-V(\theta)= {\cal S}
\eeq
where $\cal S$ is a constant.   We can see directly that the instanton corresponds to motion in minus the effective potential 
\beq
V_{eff.}(\theta)=\haf 1\frac{(l-\sigma\cos\theta)^2}{I\sin^2\theta}+V(\theta)
\eeq
that is in general divergent at both $\theta = 0$ and $\theta=\pi$.  $V(\theta)$ is assumed to be a well behaved potential with symmetric minima at the north and south poles.  We can adjust $l=\sigma$, which removes the divergence at the north pole, or $l=-\sigma$ which removes it at the south pole, but it is not possible to remove the divergence at both poles simultaneously.  We let the reader verify that for $l=\sigma$, any trajectory which starts with zero velocity at the north pole at $\tau =-\infty$, moves with infinite velocity at the south pole at a finite time.   It is easy to see that such a trajectory has infinite Euclidean action.  The boundary condition that $\theta,\dot\theta\to 0$ as $\tau\to-\infty$ implies that the integration constant ${\cal S}=0$.  Then
\bea
S_0&=&\int d\tau\lb \haf I\dot\theta^2+V_{eff.}(\theta)\rb =\int_0^\pi d\theta\sqrt{2I V_{eff.}(\theta)}\cr
&=&\int_0^\pi d\theta\sqrt{\sigma^2\tan^2\lb\theta/2\rb +2IV(\theta )}=\infty
\eea
since the integral diverges at the south pole $\theta=\pi$.  Thus $e^{-S_0/\hbar}=0$ and tunneling is suppressed for all $\sigma$, {\it i.e.} for all spins.  We see that it is crucial to consider the complex critical points of the full action in order not to get a misleading, erroneous conclusion.

{\it Quantum mechanical system - } Our result can be confirmed by looking at the corresponding Schr\"odinger quantum mechanical system, where it is easy to see, incontrovertibly, that tunneling is suppressed for all non-zero values of the spin $\sigma$.  The Lagrangian in Eq. (\ref{1}), although conceived as to describe a semiclassical spin, can be equally well thought of as the Lagrangian that determines the dynamics of a charged particle on a two-sphere that is subject to the magnetic field of a magnetic monopole located at the center of the sphere.  Evidently the electric field due to the charged particle itself is not taken into account, since it is topologically impossible to have net charge on a compact manifold such as a sphere.  Suppression of tunneling would imply that the ground state of this system is degenerate even in the presence of an easy axis, reflection symmetric potential.

The Schrödinger description of the quantum system of Eq. (\ref{2}) corresponds to the transverse (spherical) Laplacian in the presence of a gauge field of a magnetic monopole at the center of the sphere
\beq
\lb -{\hbar^2\over {2 I}}\lb\hat s\times(\vec\nabla -i\sigma\vec A^\pm)\rb^2+V(\theta )\rb\Psi^\pm=E\Psi^\pm
\eeq
where the gauge field  is actually a connexion on a non-trivial fibre bundle.   $A_\theta=0$ and the azimuthal component of the gauge field is given by
\beq
A_\varphi =\begin{cases}A_\varphi^+= \sigma (1-\cos\theta)/\sin\theta &\theta\in\left[0,\pi\right)\cr 
A_\varphi^-=-\sigma (1+\cos\theta)/\sin\theta& \theta\in\left(0,\pi\right]\end{cases}
\eeq
where $A_\varphi^+$ is related to $A_\varphi^-$ by a gauge transformation $U(\varphi)=e^{i2\sigma\varphi}$, on the domain where they are both defined, $\theta\in \lb0,\pi\rb$.  The eigensections of this problem in the absence of the potential are well studied, in the seminal paper of Wu and Yang, \cite{wy}, where the monopole harmonics are defined as sections of the associated vector bundle, and they are analog to the usual spherical harmonics (notationally, our $\sigma$ corresponds to their $q$). We can expand our solution in terms of the monopole harmonics for fixed $m$, since our problem has azimuthal symmetry
\bea
\Psi^\pm_{n,m}&=&\begin{cases}
\sum_{l\ge |m|}\psi_{n,l,m}\Theta_{n,l,m}(\theta)e^{i(m+\sigma)\varphi}&\theta\in\left[0,\pi\right)\cr
\sum_{l\ge |m|}\psi_{n,l,m}\Theta_{n,l,m}(\theta)e^{i(m-\sigma)\varphi}& \theta\in\left(0,\pi\right]
\end{cases}\cr
&\equiv&\begin{cases}
\psi_{n,m}(\theta)e^{i(m+\sigma)\varphi}&\theta\in\left[0,\pi\right)\cr
\psi_{n,m}(\theta)e^{i(m-\sigma)\varphi}& \theta\in\left(0,\pi\right]
\end{cases}
\eea
where $l=|\sigma|,|\sigma| +1,\cdots$ while $m=-l,-l+1,\cdots,l$.  Then in $\psi_{n,m}(\theta)$ the only vestige of the monopole that remains is that $m$ is a half odd integer for fermionic spin while it is an integer for bosonic spin.  $\psi_{n,m}(\theta)$ satisfies the equation
\bea
\lb-{\hbar^2\over 2 I}\lb\frac{1}{\sin\theta}\frac{\partial}{\partial\theta}\sin\theta\frac{\partial}{\partial\theta} 
+\frac{(m+\sigma\cos\theta)^2}{\sin^2\theta}\rb\rb\psi_{n,m}(\theta)\cr
\hskip-1cm+ V(\theta)\psi_{n,m}(\theta)=E_{n,m}\psi_{n,m}(\theta).\hbox{\quad}
\eea
This equation has a doubly degenerate spectrum due to the assumed reflection symmetry of the potential, $V(\theta)=V(\pi -\theta)$. The replacement $m\to -m,\theta\to\pi-\theta$ yields the same equation. Evidently for fermionic spins, all levels are doubly degenerate, {\it i.e.} especially the ground state.   For bosons, the state $m=0$ is not paired, however,  the ground state is achieved for $m=\pm\sigma$.   If  $m\ne\pm\sigma$ the spin must stay away from both poles  where the effective potential $\frac{(m+\sigma\cos\theta)^2}{\sin^2\theta}$ diverges, which forces it into regions where the potential $V(\theta)$ is non-vanishing and correspondingly lifts the energy.  For $m=\pm\sigma$  the effective potential diverges at only one of the poles, and the wave function is localized at the opposite pole, giving rise to the doubly degenerate ground states.  The ground state being doubly degenerate implies the absence of tunneling.  

{\it First order theory - }  We can consider our system when only the first order term is considered, to show that there also the tunneling is suppressed for all spins.  The corresponding Euclidean action is just obtained from Equ. (\ref{3}) by putting $I=0$
giving the corresponding equations of motion:
\bea
-\frac{\partial V(\theta)}{\partial\theta} +i\sigma\sin\theta\dot\varphi&=&0\label{78}\\
 -i\sigma\sin\theta\dot\theta&=&0\label{79}
\eea
Multiplying the first by $\dot\theta$, the second by $\dot\varphi$ and adding the two yields simply $-\frac{\partial V(\theta)}{\partial\theta} \dot\theta=0$ {\it i.e.} $V(\theta)= const.=0.$  This is just a special case of the general result, that the conserved Hamiltonian for a Lagrangian theory that is first order, is just given by the potential.  In general we find that the conserved energy is just $V(\theta,\varphi )=0$ (where we have normalized the potential so that it vanishes at the initial point, and hence always).  Thus any instanton must satisfy this constraint.  The instantons of Refs. \cite{l} and \cite{c} can be easily reproduced using this analysis.  For the easy-axis case that we study here, $V(\theta,\varphi )\to V(\theta)=0$, there simply are no solutions.  We can even allow for the possibility of solutions through complex field variables, $\theta\to\theta+i\xi$, however, the equation $V(\theta+i\xi)=0$ still affords no solutions since $V(\theta+i\xi)=u(\theta ,\xi)+iv(\theta ,\xi)=0$ requires $u(\theta ,\xi)=0$ and $v(\theta ,\xi)=0$.  But these are harmonic functions, the real and imaginary parts of a holomorphic function.  The level curves of $u(\theta ,\xi)$ are the paths of steepest descent of $v(\theta ,\xi)$ and vice versa.  It is impossible that both of them remain constant along any path.  An evident generalization to the case $V(\theta,\varphi )=V(\theta)U(\varphi)$ is left to the reader, where the same conclusions can be drawn.  Thus in the first order theory, the tunneling is suppressed for all spins for the simple reason that there are no instantons.

{\it Conclusions - } We have shown that there is suppression of quantum tunneling for both fermionic and bosonic spin systems in the case of easy-axis, azimuthally and reflection symmetric quantum spin systems.  We have also shown that it is absolutely crucial to take into account complex critical points of the Euclidean action when the Minkowski action contains $t$-odd terms.   In the present case, this leads to suppression of tunneling of macroscopic spin systems.  The experimental verification of our results should be interesting.  

{\it  Acknowledgments - } We thank The Institute of Physics, Bhubaneswar for hospitality, where this work was started, Bum-Hoon Lee, Jeong-Hyuck Park and Corneliu Sochichiu of CQUeST for useful discussions, (Paul) Hoong-Chien Lee and the Graduate Institute of Systems Biology and Bioinformatics, College of Science, National Central University, Zhongli, Taiwan for hospitality while this work was completed, Yutaka Hosotani, Department of Physics, Osaka University for hospitality where this work was written up, and  NSERC of Canada and the Center for Quantum Spacetime of Sogang University with grant number R11-2005-021,  for financial support.  
\vfill


\end{document}